\documentclass[letterpaper,twocolumn,prl,aps,superscriptaddress,amsmath,amssymb,floatfix]{revtex4-1}
\usepackage{mathptmx}
\usepackage[latin9]{inputenc}
\usepackage{color}
\usepackage{amsmath}
\usepackage{amssymb}
\usepackage{graphicx}
\usepackage{esint}
\usepackage[unicode=true,
 bookmarks=true,bookmarksnumbered=false,bookmarksopen=false,
 breaklinks=false,pdfborder={0 0 1},backref=false,colorlinks=true]
 {hyperref}
\hypersetup{
 linkcolor=magenta,urlcolor=blue,citecolor=blue,pdfstartview={FitH},hyperfootnotes=false}

\makeatletter

\usepackage{textcomp}
\usepackage{epstopdf}

\usepackage{amsfonts}

\pdfpageheight\paperheight
\pdfpagewidth\paperwidth

\@ifundefined{textcolor}{}{%
 \definecolor{BLACK}{gray}{0}
 \definecolor{WHITE}{gray}{1}
 \definecolor{RED}{rgb}{1,0,0}
 \definecolor{GREEN}{rgb}{0,1,0}
 \definecolor{BLUE}{rgb}{0,0,1}
 \definecolor{CYAN}{cmyk}{1,0,0,0}
 \definecolor{MAGENTA}{cmyk}{0,1,0,0}
 \definecolor{YELLOW}{cmyk}{0,0,1,0}
}

\usepackage{xcolor}\usepackage{soul}
\newcommand{\ket}[1]{\ensuremath{\left|#1\right\rangle}}

\definecolor{blue}{rgb}{0,0,1}
\definecolor{red}{rgb}{1,0,0}
\definecolor{green}{rgb}{0,1,0}

\makeatother

\begin{document}

\title{Perturbation impact of spectators on a cross-resonance gate in a tunable coupling superconducting circuit}

\author{T.-Q. Cai}
\thanks{These two authors contributed equally to this work.}
\affiliation{Center for Quantum Information, Institute for Interdisciplinary Information Sciences, Tsinghua University, Beijing 100084, China}

\author{X.-Y. Han}
\thanks{These two authors contributed equally to this work.}
\affiliation{Center for Quantum Information, Institute for Interdisciplinary Information Sciences, Tsinghua University, Beijing 100084, China}

\author{Y.-K. Wu}
\affiliation{Center for Quantum Information, Institute for Interdisciplinary Information Sciences, Tsinghua University, Beijing 100084, China}

\author{Y.-L. Ma}
\affiliation{Center for Quantum Information, Institute for Interdisciplinary Information Sciences, Tsinghua University, Beijing 100084, China}

\author{J.-H. Wang}
\affiliation{Center for Quantum Information, Institute for Interdisciplinary Information Sciences, Tsinghua University, Beijing 100084, China}

\author{Z.-L. Wang}
\affiliation{Center for Quantum Information, Institute for Interdisciplinary Information Sciences, Tsinghua University, Beijing 100084, China}

\author{H.-Y Zhang}
\affiliation{Center for Quantum Information, Institute for Interdisciplinary Information Sciences, Tsinghua University, Beijing 100084, China}

\author{H.-Y Wang}
\affiliation{Center for Quantum Information, Institute for Interdisciplinary Information Sciences, Tsinghua University, Beijing 100084, China}

\author{Y.-P. Song}\email{ypsong@mail.tsinghua.edu.cn}
\affiliation{Center for Quantum Information, Institute for Interdisciplinary Information Sciences, Tsinghua University, Beijing 100084, China}

\author{L.-M. Duan}\email{lmduan@tsinghua.edu.cn}
\affiliation{Center for Quantum Information, Institute for Interdisciplinary Information Sciences, Tsinghua University, Beijing 100084, China}

\begin{abstract}
Cross-resonance (CR) gate has emerged as a promising scheme for fault-tolerant quantum computation with fixed-frequency qubits. We experimentally implement entangling CR gate by using a microwave-only control in a tunable coupling superconducting circuit, where the tunable coupler provides extra degrees of freedom to verify optimal condition for constructing CR gate. By developing three-qubit CR Hamiltonian tomography protocol, we systematically investigate the dependency of gate fidelities on spurious qubit interactions and present the first experimental approach to the evaluation of the perturbation impact arising from spectator qubits. Our results reveal that the spectator qubits lead to reductions in CR gate fidelity dependent on $ZZ$ interaction and particular frequency detunings between spectator and gate qubits, demonstrating a more serious impact from the target spectator than from the control spectator. Our experiments uncover optimal CR operation regime and provide insight into an improvement of the CR gate by suppression of unwanted qubit interactions.
\end{abstract}

\maketitle

\section{Introduction}

As enormous progress in superconducting quantum processors has been made towards more complex networks of qubits, it becomes increasingly crucial to develop robust protocols for multi-qubit control~\cite{Gambetta_2017,song2019generation,houck2012chip,neill2018a}. In complex superconducting circuits with larger numbers of qubits, the fidelity of quantum algorithms begins to be dominated by unwanted qubit interactions, increased decoherence, and frequency-crowding, all inherent to traditional frequency-tuned architectures~\cite{Pinto_2010}. Alternatively, a microwave-only control scheme can provide frequency-selectivity and allow to use fixed-frequency computational qubits, thereby minimizing the sensitivity of the qubits with respect to the sources of possible noise~\cite{poletto2012entanglement,chow2013microwave}. Cross-resonance (CR) gate is an entangling gate for fixed-frequency qubits by using only microwave control. To perform a CR gate, a microwave drive is applied on a control qubit at a frequency of a target qubit~\cite{chow2011simple,Rigetti_2010,pommerening2017multiqubit,sheldon2016procedure,ware2015flux,Kirchhoff_2018}. The gate itself obtains a Clifford group operator $[ZX]^{1/2}$, which is locally equivalent to a universal gate CNOT by only one additional local rotation of each qubit.~\cite{Rigetti_2010,pommerening2017multiqubit}. Consequently, the CR gate scheme has a strong appeal to the multi-qubit control in superconducting architectures using fixed-frequency transmon qubits, thus allowing the qubits to be operated at their optimal bias points for coherence; also, it only requires a single microwave drive line for applying the drive tone to the control qubit and, thereby efficiently reduces the circuit complexity.

The CR gate has been demonstrated, obtaining a quantum process tomography (QPT) gate fidelity of 81\%~\cite{chow2011simple}. The gate process has been improved via a calibration procedure, achieving an interleaved randomized benchmarking fidelity over 99\%~\cite{sheldon2016procedure}. A fast two-qubit gate relies on a large coupling, but leading to crosstalk between qubits. Accordingly, the CR implementation is hindered by the trade-off between a long gate time and a high gate fidelity. Recent study has revealed that, besides the cross resonance component $ZX$, the CR drive Hamiltonian also involves other unwanted qubit interactions, such as $IX$, $IY$, $IZ$, $ZZ$ term, etc~\cite{sheldon2016procedure}. Particularly, CR gate has been used in small-scale multi-qubit demonstrations of fault-tolerant protocols~\cite{C_rcoles_2015,Gambetta_2017}. In practice, however, computational qubits of CR gate can not be efficiently isolated from environment and are inevitably exposed to neighboring qubits owing to mutual interactions in a quantum processor~\cite{arute2019quantum,blok2020quantum}. A recent theoretic study on CR gate reveals detrimental multi-qubit frequency collisions as a control or target qubit couples to a third spectator qubit~\cite{Malekakhlagh_2020}, thus leading to reduction in gate fidelity. To eliminate this deadly impact, it becomes crucial to study the dependency of the unwanted components on the coupling between the qubits. In particular, it remains desired for an experimental investigation of optimal CR gate operation regime in the presence of spectator qubits. These key problems, however, have not yet been solved due to a less control of the interactions between the qubits. Fortunately, experimental realization of tunable couplers provides a way to adjust qubit interactions, and hence offers a possibility for mitigating unwanted couplings~\cite{Mundada_2019,Li_2020,Han_2020,Bialczak_2011,Chen_2014,McKay_2016}. 

In this work, with exploiting flux-controlled tunable couplers, we address these crucial barriers to optimizing CR gate control by systematically investigating the dependency of gate fidelities on spurious interaction components. This study presents the first experimental approach to the evaluation of the perturbation impact arising from the spectator qubits, providing a guiding principle to improve the CR gate fidelity by suppression of the qubit-spectator interactions. Our experimental results reveal that the spectator qubits have a significant impact on the computational gate qubits, leading to reduction in the gate fidelity dependent on the frequency resonance poles and the induced $ZZ$ interaction between the spectator and the gate qubits. By optimally tuning the inter-qubit detuning and flux bias on the coupler, we achieve a CR gate fidelity of 98.5\%, primarily limited by qubit decoherence. 

\section{Results}
\subsection{Isolated Cross-resonance Gate}

\begin{figure}[t]
\includegraphics{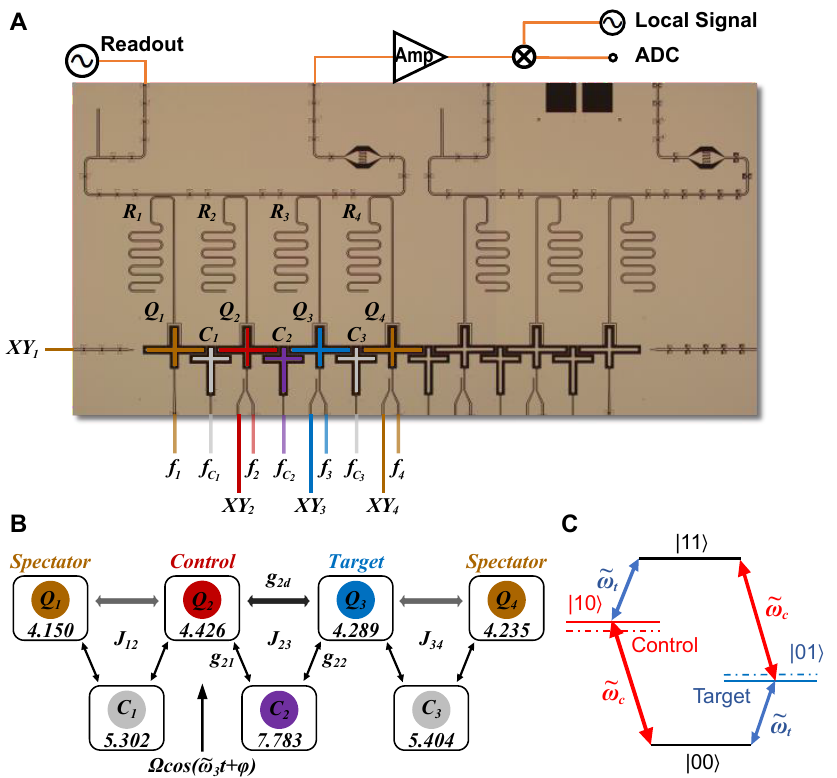}
\caption{(A) Optical micrograph of thirteen Xmon qubits with the bottom six qubits serving as tunable couplers. Each computational qubit on both sides of the couplers has independent $XY$ and $Z$ control, and is coupled to a separate $\lambda/4$ resonator for simultaneous and individual readout. The couplers each have an individual flux-bias line for frequency tunability. The combination of direct capacitive coupling and indirect tunable coupling via the coupler allows us to continuously vary adjacent qubit coupling from positive to negative. The qubits and couplers in false colors demonstrate the subsystem used in the experiment. (B) Sketch of the the subsystem. The red, blue, purple and yellow circles represent the control, target, coupler and spectator qubits, respectively. The qubit frequencies (in GHz) are displayed in the corresponding boxes. The two computational qubits are coupled with an effective coupling $J_{23}$ via a direct coupling with strength $g_{2d}$ and indirect couplings through the coupler with $g_{21}$ and $g_{22}$, respectively. (C) Energy spectrum of a pair of weakly coupled qubits as $J_{23}$ is small compared to $\Delta(\phi)$. Dashed (solid) lines depict uncoupled (coupled) energy levels for $Q_2$ (red, labeled as control) and $Q_3$ (blue, labeled as target).}
\label{fig:Fig1}
\end{figure}

Our quantum processor consists of seven transmon qubits ($Q_i, \, i=1 \sim 7$) with each pair of neighboring qubits mediated via a frequency-tunable coupler ($C_j, \, j=1 \sim 6$), as shown in Fig.~\ref{fig:Fig1}(A). Each tunable coupler has a symmetric Josephson junction with a cross-shaped capacitor sandwiched between two neighboring qubits, contributing to the total coupling between two computational qubits. Each qubit, $Q_i$ and $C_j$, has a dedicated flux bias line to tune its frequency by threading a magnetic flux through transmon junction loop. In our experiments, the qubits, $Q_i \, (i=2, \, 3)$ and $Q_i \, (i=1, \, 4)$ as outlined in Fig.~\ref{fig:Fig1}(B), are used to implement the CR gate as the computational gate qubits and spectator qubits, respectively. 

The two gate qubits $Q_i \, (i=2, \, 3)$ each couple to the tunable coupler $C_2$ with a coupling strength $g_{21},\, g_{22}$, as well as to each other with a direct capacitive coupling strength $g_{2d}$. Both qubits are negatively detuned from the coupler, $\Delta_i$($\phi$)=$\omega_i-\omega_c<0 \, (i=2, \, 3)$, where $\omega_{2,3}$, $\omega_{c}$ are the frequencies of $Q_2$, $Q_3$ and $C_2$, respectively.  The experimentally extracted parameters, $g_{21}, \, g_{22}/2\pi = 63$ MHz, $ g_{2d}/2\pi \sim 5.5$ MHz, give a dispersive coupling, $g_{21},\, g_{22} \ll \left\vert{\Delta_i(\phi)}\right\vert$ (see Supplementary Material~\cite{Supplementary} for details). We apply a cross-resonance (CR) drive pulse on the control qubit $Q_2$, $\Omega\cos(\omega_dt+\phi)$, with an amplitude $\Omega$, frequency $\omega_d$ and phase $\phi$. When the qubit drive is present, the system Hamiltonian states as,
\begin{eqnarray}\label{Eq1}
\begin{split}
H/\hbar&=\sum_{i=2, 3}\frac{1}{2}{\widetilde{\omega}_i}\sigma^z_i+J_{23}(\sigma^+_2\sigma^-_3+\sigma^+_3\sigma^-_2)\\
&+\Omega\cos(\omega_dt+\phi)\sigma_2^x,
\end{split}
\end{eqnarray}
where $\sigma_\alpha^{x}$, $\sigma_\alpha^{z}$, $\sigma_\alpha^{+}$, $\sigma_\alpha^{-}$ ($\alpha=2,\, 3$) are the Pauli $X$, Pauli $Z$, raising and lowering operators for $Q_2$ and $Q_3$ respectively; $\widetilde{\omega}_2=\omega_2+\frac{J_{23}}{\Delta(\phi)}$, $\widetilde{\omega}_3=\omega_3-\frac{J_{23}}{\Delta(\phi)}$, $J_{23}=g_{2d}+\frac{g_{21} g_{22}}{\Delta(\phi)}$, $\frac{1}{\Delta(\phi)}=(\frac{1}{\Delta_2(\phi)}+\frac{1}{\Delta_3(\phi)})/2$~\cite{blais2007quantum,bravyi2011schrieffer,yan2018tunable}. The combination of two terms, $g_{2d}+\frac{g_{21} g_{22}}{\Delta(\phi)}$, gives the total effective qubit-qubit coupling $J_{23}$, which can be adjusted by varying the coupler frequency through $\Delta(\phi)$. Energy spectrum shown in Fig.~\ref{fig:Fig1}(C) depicts the corresponding qubit frequency shift as $J_{23}$ is small compared to $\Delta(\phi)$. Since the tunability is continuous, one can always find a critical value to turn off the effective coupling $J_{23}$, as well as the static $ZZ$ coupling~\cite{Li_2020}. On the condition that $\Omega, \,J_{23} \ll \Delta(\phi)$, and the drive frequency $\omega_d$ is in resonance with the target qubit ($Q_3$) frequency $\widetilde{\omega}_3$, under the consideration of crosstalks on the processor chip and off-resonance drive on the control qubit, the effective drive Hamiltonian can be expressed as $H_{eff}/\hbar=u_1 ZX + u_2 ZY + u_3 ZZ + u_4 ZI + u_5 IX + u_6 IY + u_7 IZ$~\cite{Magesan_2020,sheldon2016procedure}. The first one is the cross-resonance (CR) term, while the rests are the unwanted residual qubit interaction terms in the gate operation. For instance, the forth term represents an ac-Stark shift due to the off-resonance drive on the control qubit, and the fifth one reflects the crosstalk on the target qubit. Based on energy-basis representation method, in consideration of both higher-level effects of qubits and classical crosstalks, the CR term can be calculated by $u_1 = J_{23} \Omega (1-A_c)(\frac{\nu_{3,01} \nu^2_{2,12}}{\Delta_{23}+\alpha_2}-\frac{2\nu_{3,01} \nu^2_{2,01}}{\Delta_{23}})$~\cite{Malekakhlagh_2020}, where $A_c$ denotes a suppression in the drive tone on the control qubit, $\nu_{i, 01}, \, \nu_{i, 12} \,(i=2, \, 3)$ are extracted dimensionless parameters which are defined in Supplementary Materials~\cite{Supplementary}, $\alpha_2$ is anharmonicity of $Q_2$, $\Delta_{23}=\widetilde{\omega}_2 - \widetilde{\omega}_3$.

\begin{figure*}[t]
\includegraphics{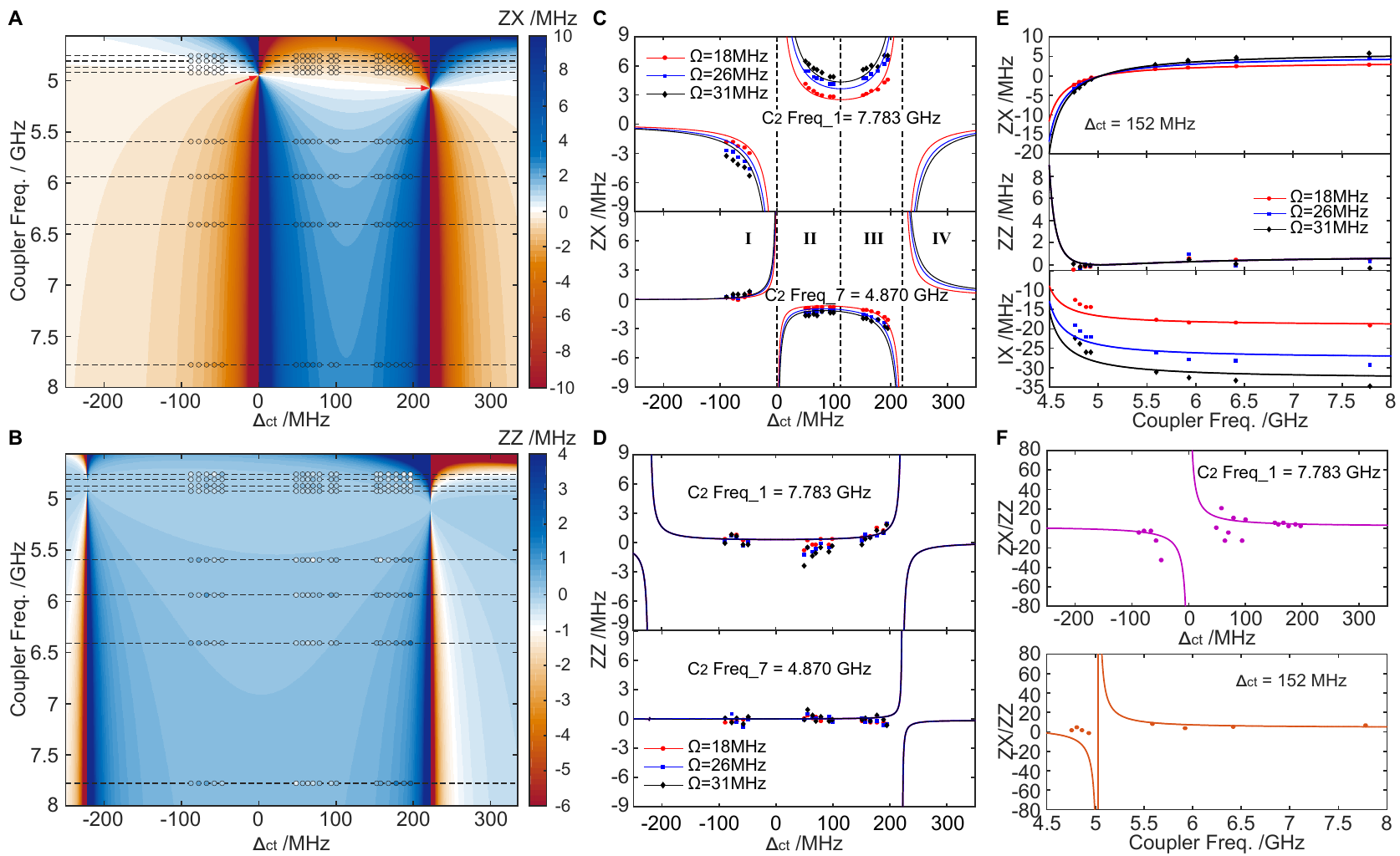}
\caption{(A),(B) Numerical simulations and measurements of $ZX$ and $ZZ$ interactions with the fixed CR drive amplitude $\Omega=18$ MHz. The measured data points are displayed in black circles with color intensity inside to identify the interaction strength. (C),(D) Cross-sectional views of (A) and (B) at two representative coupler frequencies. The three sets of data present the $ZX$ and $ZZ$ interactions with respect to the CR drive amplitude of $18$, $26$ and $31$ MHz, respectively. The large $ZX$ rate while small $ZZ$ interaction in region III defines the optimal operating regime for the CR gate. (E) Three interaction components of $ZX$, $ZZ$ and $IX$ vary as a function of the coupler ($C_2$) frequency in a range of $4.5 \sim 8.0$ GHz (calculation data) with three different CR drive amplitudes. (F) The experimental (dots) and numerical (solid lines) results of $ZX/ZZ$ ratio vary with the qubit frequency detuning $\Delta_{ct}$ (top panel) and the coupler ($C_2$) frequency (bottom panel). The optimal operating condition is expected in the region where the $ZX/ZZ$ ratio is large but insensitive to the coupler frequency.}
\label{fig:Fig2}
\end{figure*}

\begin{figure*}[t]
\includegraphics{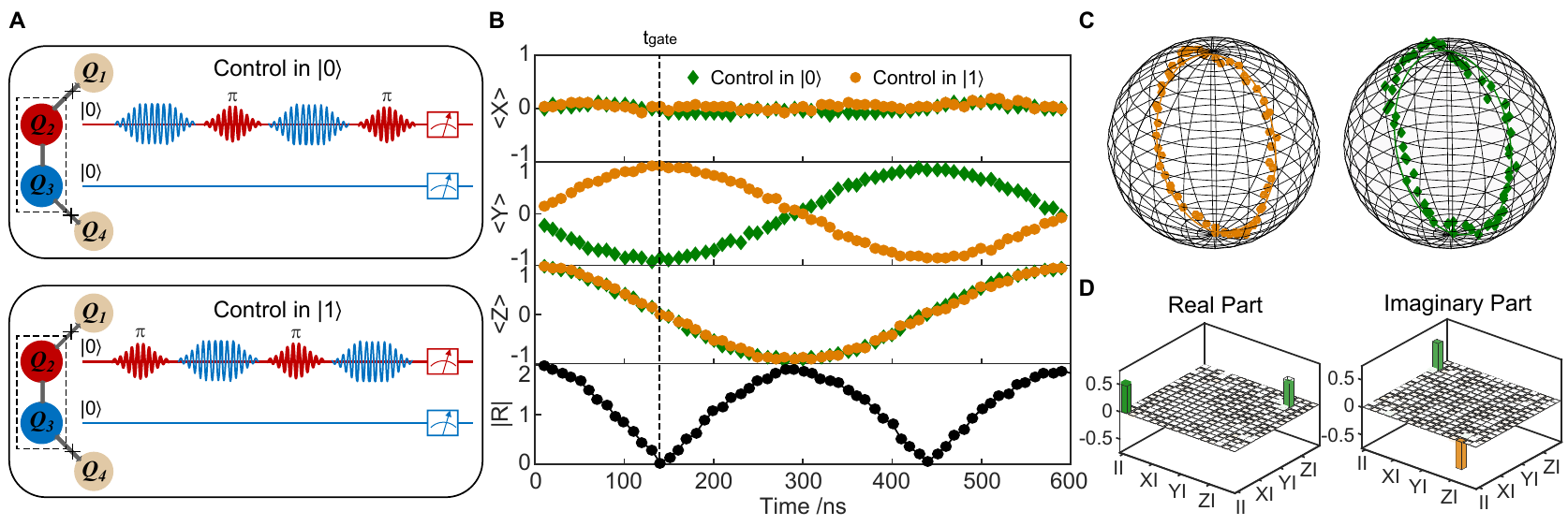}
\caption{(A) Echoed pulse sequence involves a $\pi$ pulse and a change of sign of the CR/2 drive on the control qubit. (B) CR Rabi oscillations on the target qubit projected onto $x$, $y$, and $z$ for the control in $\ket{0}$ and $\ket{1}$. Sinusoidal oscillations are expected for a $ZX$ drive. $R$ vector at 0 indicates the gate time of maximal entanglement. (C) Trajectory of the target qubit state on the Bloch sphere during the echoed CR gate evolution for the control in $\ket{0}$ (left column) and $\ket{1}$ (right column). A perfect $ZX$ gate would result in a circle on the surface of the Bloch sphere. (D) Measured QPT result for an entangling CR gate with a gate length of $220$ ns (CR pulse: 140 ns; two single-qubit $\pi$ pulses: 40ns for each). The solid black outlines are for the ideal gate.}
\label{fig:Fig3}
\end{figure*}

To verify optimal implementation parameters and extract error terms in the gate operation, we numerically calculate CR Hamiltonian components based on the lowest-order energy-basis representation method~\cite{Malekakhlagh_2020} with experimental parameters, and we plot two primary interaction terms, $ZX$ and $ZZ$, as a function of control-target qubit frequency detuning $\Delta_{ct}$ and coupler frequency, as shown in Fig.~\ref{fig:Fig2}(A) and (B). The results reveal that the interaction components are sensitive to the frequency detuning, featuring two-qubit resonance poles as the detuning crosses the gate parameters $\Delta_{ct}=0$, $\Delta_{ct}=\pm \alpha_i=\pm 222$ MHz ($i=2, \,3$), and thus divide the gate operation into the distinct regions labeled with I, II, III and IV. Moreover, the interaction terms undergo the turning points indicated by red arrows, slightly dependent on the frequency detuning, as the coupler frequency passes across the transition point. We experimentally measure the CR Hamiltonian and fit Rabi oscillations with a Bloch equation model function~\cite{sheldon2016procedure}, with the used pulse sequence sketched in Supplementary Materials~\cite{Supplementary}. The CR drive Hamiltonian can then be derived in terms of the six possible interactions $IX$, $IY$, $IZ$, $ZX$, $ZY$ and $ZZ$. The measured $ZX$ and $ZZ$ interactions at eight different coupler frequencies, as shown in Fig.~\ref{fig:Fig2}(A) and (B), are positionly illustrated in black circles with color intensity inside, which are consistent with the numerical calculations. To highlight their dependence on the frequency detuning, we plot the measured interactions of $ZX$ and $ZZ$ (dots) and the simulations (solid lines) in Fig.~\ref{fig:Fig2}(C) and (D) for two representative coupler frequencies with different CR drive amplitudes. The aformentioned distinct regions are clearly distinguished with the detuning transitions. Furthermore, we selectively plot three interaction components of both the measured and the calculated $ZX$, $ZZ$ and $IX$ in Fig.~\ref{fig:Fig2}(E) as a function of the coupler frequency with three different drive amplitudes and a fixed $\Delta_{ct}=152$ MHz. With the increase of the DC flux bias, the coupler qubit frequency is reduced to be close to the gate qubit frequency. Consequently, both the experimental and the simulated $ZX$ interactions slowly vary from positive to negative but turn to a rapid change as the coupler frequency goes below the turning point around $4.6$ GHz for each drive amplitude, demonstrating a tunability range of about $5.6 \sim -3.9$ MHz (experimental data) with $\Omega = 31$ MHz as an example. The calculated $ZZ$ term shows a relatively smaller variation range, keeping positive interaction but increasing rapidly below the turning point. The $IX$ interaction, however, monotonically declines as the coupler frequency decreases. Compared with the $ZX$ or $ZZ$ interaction, the $IX$ term shows a much stronger dependence on the CR drive. We find that the large $ZX$ rate while relatively small static $ZZ$ interaction in region III defines an optimal operating regime in our experiment, which is confirmed by the experimental data and numerical calculations (solid lines) shown in the top panel of Fig.~\ref{fig:Fig2}(F). In addition, the bottom panel of Fig.~\ref{fig:Fig2}(F) implies that $ZX/ZZ$ is less sensitive to the coupler frequency except for the region near $ZZ = 0$.

To suppress the unwanted CR components, we verify an appropriate CR drive phase at which  the $ZX$ component is maximized whereas the $ZY$ is zero by measuring the CR Hamiltonian parameters as a function of the drive phase, as shown in Supplementary Materials~\cite{Supplementary}. To further eliminate the cross-talk term $IX$ and other unwanted interactions $ZZ$ and $ZI$, we perform the CR Rabi experiment using an echo scheme to refocus these terms~\cite{sheldon2016procedure}. The echo sequence involves a $\pi$ pulse sandwiched between two half-length CR drives (CR/2) with reversed polarity on the control qubit, as sketched in Fig.~\ref{fig:Fig3}(A). Clearly, in Fig.~\ref{fig:Fig3}(B), the echo scheme indeed improves the CR Rabi oscillations on the target qubit, which are much closer to sinusoidal oscillations expected for a $ZX$ drive compared with the one obtained by using the regular pulse sequence without echo integration. The $R$ vector plotted in the figure is a norm distance between Bloch vectors of the target qubit corresponding to the control qubit in $\left|0\right\rangle$ and $\left|1\right\rangle$, which can be defined as $||\overrightarrow{R}||=\sqrt{\sum_{r=x,y,z}(\left\langle{r_0}\right\rangle+\left\langle{r_1}\right\rangle)^2}$. The two qubits are maximally entangled when $R$ vector goes to zero, and hence we can use $||R||$ to estimate the gate length for performing the entangling gate, which is about 140 ns as marked in Fig.~\ref{fig:Fig3}(B). Apparently, the result shows a full contrast of $\left\langle{y}\right\rangle$ while non-oscillation of $\left\langle{x}\right\rangle$, indicating that the $IY$ term has been cancelled out. As illustrated in Fig.~\ref{fig:Fig3}(C), the trajectory of the target qubit is depicted in a Bloch sphere during the CR gate operation, showing a near perfect circle on the surface of the Bloch sphere, which further confirms that the echo scheme improves the gate evolution. 

The ratio of the desired $ZX$ term to the unwanted $ZZ$ component uncovers optimal detuning to achieve low coherent error for the CR gate. According to the results shown in Fig.~\ref{fig:Fig2}(F), we finally choose $\Delta_{ct}=137$ MHz with the coupler ($C_2$) frequency of $7.783$ MHz to be the optimal operating position where one expects to yield a high gate fidelity and low two-qubit coherent error. Indeed, we verify the CR gate fidelity by varying the inter-qubit detuning and DC flux bias on the coupler. We perform quantum process tomography (QPT) for the CR gate by implementing 16 independent two-qubit input states and construct the Pauli transfer matrix $\chi$. The gate fidelity can be determined from the $\chi$ matrix through the expression $F = tr(\chi_{\mathrm{exp}}\chi_{\mathrm{ideal}})$, where $\chi_{\mathrm{exp}}$ and $\chi_{\mathrm{ideal}}$ are the experimental and ideal $\chi$ matrix. Fig.~\ref{fig:Fig3}(D) shows the $\chi_{\mathrm{exp}}$ and $\chi_{\mathrm{ideal}}$ for the CR entangling gate acquired under the optimal condition. The gate fidelity from QPT is 98.5\% from maximum-likelihood estimation, primarily limited by qubit decoherence~\cite{korotkov2013error}.

 \begin{figure*}[t]
\includegraphics{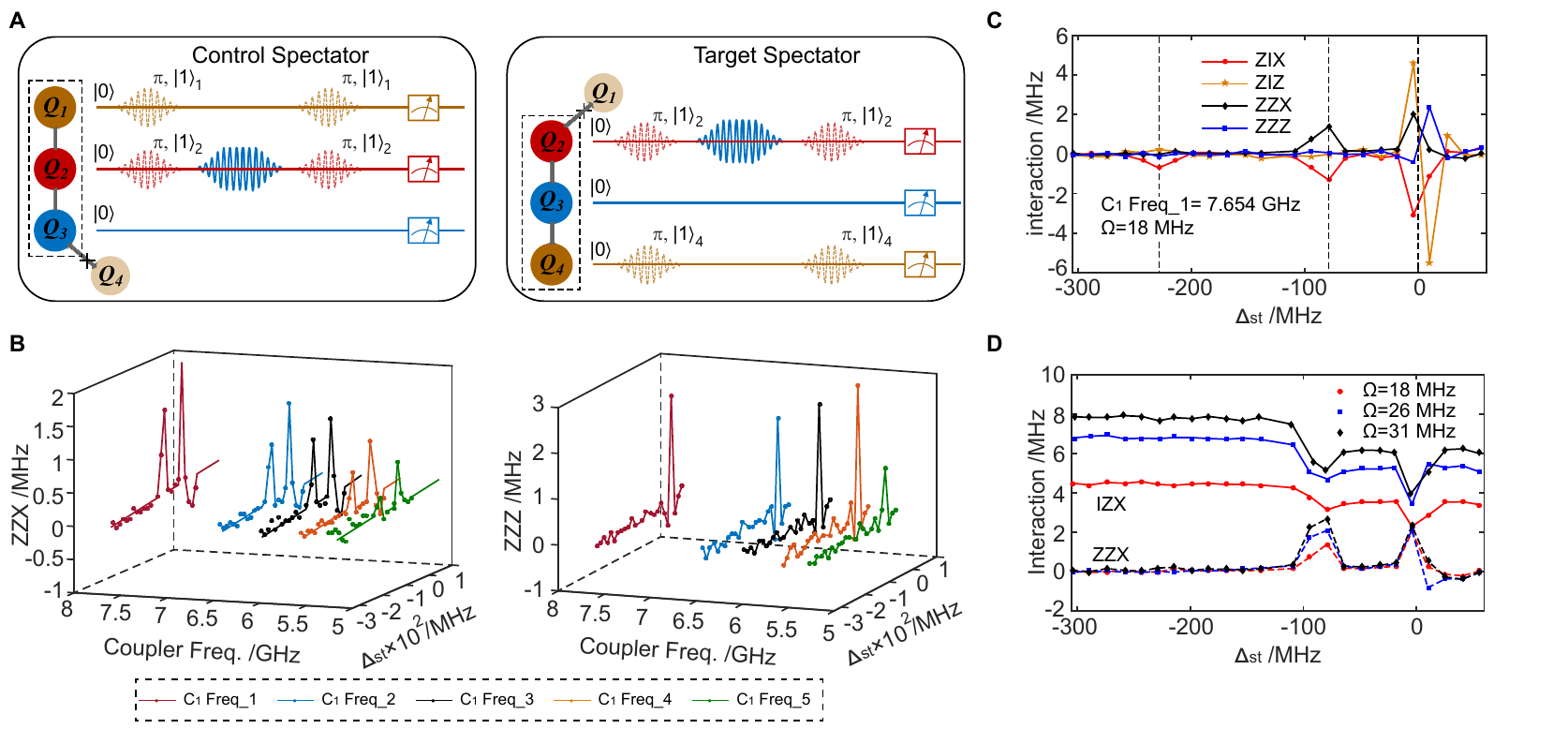}
\caption{(A) Schematic pulse sequences for measuring three-qubit Hamiltonian tomography with either a control spectator qubit (left panel) or a target spectator qubit (right panel). For the control-spectator case, a CR pulse is applied on the control qubit in the spectator-control subspace of $\ket{00},\, \ket{01},\, \ket{10},\, \ket{11}$, respectively. For example, when measuring Rabi oscillation of the target qubit in the subspace of $\ket{10}$ and $\ket{11}$, two $\pi$ pulses are consequently applied on $Q_1$ before and after the CR pulse. Three-qubit simultaneous single-shot readout is then performed to measure the interaction terms based on Eq.~(\ref{Eq4}). Note that the coupling between $Q_3$ and $Q_4$ is always closed during the process. Similarly, for the target spectator case, the CR pulse is applied on the control qubit in the target-spectator subspace with $\ket{00},\, \ket{01},\, \ket{10}$ and $\ket{11}$ respectively, keeping the coupling between $Q_1$ and $Q_2$ closed. (B) The dominant three-qubit CR Hamiltonian $ZZX$ ($ZZZ$) term in the control spectator case varying with the change of $\Delta{st}$ (tuning qubit frequency of $Q_1$) and $J_{12}$ (changing coupler frequency of $C_1$). (C) The dominant interaction terms, in the control spectator case, varying as a function of $\Delta_{st}$ with a fixed coupler frequency of $C_1$ at $7.654$ GHz and a CR drive amplitude at $18$ MHz. Apparently, all interaction terms demonstrate extreme changes in certain detuning regions ($\Delta_{st}=0,\, -85,\, -222$ MHz), revealing unwanted energy excitations happened in the CR gate operation. (D) Three-qubit $IZX$ and $ZZX$ terms vary with the $\Delta_{st}$ and CR drive amplitude in the control spectator case. The $IZX$ is crucial for implementing CR gate, while the $ZZX$ is the error term generated from the spectator qubit. Clearly, the interaction terms are more sensitive at the specific detuning positions with larger drive amplitude.}
\label{fig:Fig4}
\end{figure*}

\subsection{Cross-resonance Gate with Spectator Qubits}
Since the coupling with surrounding qubits inevitably exists in multi-qubit system, the dispersive coupling between gate qubits and spectator qubits will undoubtedly affect the gate fidelity. In order to identify a more realistic scenario of gate operation, we conduct CR gate with considering a third spectator qubit which could be coupled to either the control or the target qubit, see the schematic circuit in Fig.~\ref{fig:Fig1}(B). $Q_1$ and $Q_4$, served as control spectator and target spectator, couple to the control qubit $Q_2$ and the target qubit $Q_3$ with an effective coupling strength $J_{12}$ and $J_{34}$, respectively. Under the consideration of anharmonicity for each qubit, the system can be described by a Hamiltonian $H = H_{\mathrm{gate}} + H_{\mathrm{spec}}$, where:
\begin{equation}\label{Eq2}
\begin{split}
H_{\mathrm{gate}}/\hbar&=\sum_{i=2, 3} \omega_i a_i^\dag a_i - \frac{\alpha_i}{2} a_i^\dag a_i^\dag a_i a_i \\ 
&+J_{23} (a_2^\dag a_3 + a_2 a_3^\dag) + \Omega cos(\omega_d t + \phi) (a_2 + a_2^\dag), \\
H_{\mathrm{spec}}/\hbar&=\sum_{i=1, 4} \omega_i a_i^\dag a_i - \frac{\alpha_i}{2} a_i^\dag a_i^\dag a_i a_i \\
&+ J_{12} (a_1^\dag a_2 + a_1 a_2^\dag) + J_{34} (a_3^\dag a_4 + a_3 a_4^\dag).
\end{split}
\end{equation}
where $\omega_i, \,\alpha_i \,(i=1 \sim 4)$ are the frequencies and anharmonicities of each qubit. $H_{\mathrm{gate}}$ generates the effective two-qubit $ZX$ term as discussed above, while $H_{\mathrm{spec}}$ represents the potential effects of the spectator qubits on the gate qubits. In the dispersive regime, the coupling between gate qubits and spectator qubits results in a parasitic $ZZ$ crosstalk between $Q_1$ and $Q_2$ as well as between $Q_3$ and $Q_4$. In addition, as the spectator frequency crosses some specific values, unexpected multi-qubit resonances may induce a failure of the CR gate operation~\cite{Malekakhlagh_2020}.

Hamiltoninan tomography is a useful tool for distinguishing various interaction components and determining specific error terms in gate operation. Nevertheless, when a larger network of qubits is under consideration, the original Hamiltonian parameters should be modified to account for the impact of spectator qubits. It is obvious that the CR drive pulse is only resonant with the effective target qubit frequency, and hence, apart from the target qubit containing all the Pauli interactions $\{ I,\, X,\, Y,\, Z \}$ with others, both the control qubit and spectator qubits should only involve $\{I,\, Z \}$ interactions. Based on these, without loss of generality, we consider a minimal extension of the original isolated CR model and develop an approach for three-qubit CR Hamiltonian tomography with either a control or a target spectator qubit. 

For clarity, gate operators with a control spectator qubit are naturally formed as $\ket{spectator} \otimes \ket{control} \otimes \ket{target}=\{ I,\, Z \} \otimes \{ I,\, Z \} \otimes \{ I,\, X,\, Y,\, Z \}$. Therefore, the full CR drive Hamiltonian with the control spectator qubit has a following form:
\begin{equation}\label{Eq3}
\begin{split}
H = \frac{I \otimes I \otimes A}{2} + \frac{I \otimes Z \otimes B}{2} + \frac{Z \otimes I \otimes C}{2} + \frac{Z \otimes Z \otimes D}{2},
\end{split}
\end{equation}
where $A,\, B,\, C,\, D \in \{X,\, Y,\, Z \}$. We experimentally measure the three-qubit Hamiltonian tomography to extract primary interaction terms. This is accomplished by turning on the CR drive for some time and then measuring the Rabi oscillations on the target qubit in the $spectator \otimes control$ subspace of $\ket{00},\, \ket{01},\, \ket{10},\, \ket{11}$, for projecting the target qubit state onto $x$, $y$ and $z$ axis. Similar to the approach developed for the two-qubit Hamiltonian tomography, the Rabi oscillations can be fitted with a Bloch equation model function, $\dot{\overrightarrow {r}}_{\{00, 01, 10, 11\}}(t)= e^{Gt}\overrightarrow{r}_{\{00, 01, 10, 11\}}(0),\,(r=x,\, y,\, z)$. $\overrightarrow{r}(t)$ is the vector composed of three projecting measurement values, $\left\langle x(t)\right\rangle$, $\left\langle y(t)\right\rangle$, $\left\langle z(t)\right\rangle$, as a function of the length of the Rabi drive. G is a matrix defined as 
\begin{equation}\label{Eq4}       
\left(                 
\begin{array}{ccc}   
0  &  \Delta^{\{00, 01, 10, 11\}}  &  \Omega_y^{\{00, 01, 10, 11\}}\\  
-\Delta^{\{00, 01, 10, 11\}}  & 0 & -\Omega_x^{\{00, 01, 10, 11\}}\\ 
-\Omega_y^{\{00, 01, 10, 11\}} & -\Omega_x^{\{00, 01, 10, 11\}} & 0 \\
\end{array}
\right)                 
\end{equation}
where $\Delta^{\{00, 01,10, 11\}}$ is the control drive detuning, and $\Omega_{x,y}^{\{00, 01, 10, 11\}}$ is the Rabi drive amplitude of the $x,\, y$ component, in respect of the $spectator \otimes control$ subspace in $\ket{00},\, \ket{01},\, \ket{10},\, \ket{11}$. Accordingly, all the interaction terms in Eq.~(\ref{Eq3}) can be readily acquired. For example, $IZX = (\Omega_x^{00} - \Omega_x^{01} + \Omega_x^{10} - \Omega_x^{11})/8,\, ZZX =  (\Omega_x^{00} - \Omega_x^{01} - \Omega_x^{10} + \Omega_x^{11})/8$~\cite{Supplementary}.

The dominant effects of spectator qubits can be manifested in two aspects: the frequency shift of gate qubit due to its $ZZ$ interaction with the spectator qubits and the specific frequency detunings where two- or three- qubit resonances involve. Therefore, to further detect these effects, we separately extract the three-qubit CR drive Hamiltonian terms with changing the coupling strength $J_{12}$ between $Q_1$ and $Q_2$ and the frequency detuning $\Delta_{st}$ between $Q_1$ and $Q_3$. The schematic pulse sequence is outlined in the left panel of Fig.~\ref{fig:Fig4}(A). Here, we fix the gate qubits $Q_2$ and $Q_3$ at the optimal gate position, with $\Delta_{23}=137$ MHz, according to the isolated two-qubit CR Hamiltonian tomography shown in Fig.~\ref{fig:Fig2}. The coupler $C_2$ is biased at $7.783$ GHz, offering a positive coupling between the gate qubits.

We vary $\Delta_{st}$ and $J_{12}$ by changing the frequency of the spectator qubit $Q_1$ and the coupler $C_1$, respectively. Fig.~\ref{fig:Fig4}(B) shows the dominant three-qubit gate parameters, $ZZX$ and $ZZZ$, extracted from the Bloch equation Eq.~(\ref{Eq4}), as a function of $\Delta_{st}$ and $J_{12}$. A further cross-sectional view in Fig.~\ref{fig:Fig4}(C) displays more primary interaction terms. Apparently, in certain resonance regions, unwanted energy excitations appear, breaking down the CR gate regime. For instance, the condition of $\Delta_{st}=0$ leads to a resonance between $\ket{100}$ and $\ket{001}$, while the parameters in the region around $\Delta_{st}=-85$ MHz result in a resonance of $\ket{110}$ and $\ket{020}$. Except for these regions, the interaction terms remain almost intact with different coupling strengths. Moreover, the interaction terms, such as $ZIX$, $ZZX$, $ZIZ$ and $ZZZ$, describe the effective mediated interaction between the control spectator $Q_1$ and the target qubit $Q_3$ through the control qubit $Q_2$. These terms affect the evolution of the target qubit, and thus degrade the CR gate fidelity. Consequently, we choose an appropriate frequency of the control spectator qubit $Q_1$ so that the detuning $\Delta_{st}$ is tuned away from the resonance poles. Furthermore, these three-qubit Hamiltonian interaction terms also have a dependence on the CR drive amplitude. As an example, the $ZZX$ and $IZX$ interactions, illustrated in Fig.~\ref{fig:Fig4}(D), enhance with the increase of CR drive amplitude, which is more pronounced in the resonance pole region for the $ZZX$ term.

Similar to the case with the control spectator qubit, the gate operators with a target spectator qubit can be defined as $\ket{control} \otimes \ket{target} \otimes \ket{spectator}=\{ I,\, Z \} \otimes \{ I,\, X,\, Y,\, Z \} \otimes \{ I,\, Z \}$. Thus, the CR drive Hamiltonian with the target spectator qubit can be expressed as:
\begin{equation}\label{Eq5}
\begin{split}
H = \frac{I \otimes A \otimes I}{2} + \frac{Z \otimes B \otimes I}{2} + \frac{I \otimes C \otimes Z}{2} + \frac{Z \otimes D \otimes Z}{2},
\end{split}
\end{equation}
Naturally, based on the Bloch equation Eq.~(\ref{Eq4}), we can conduct three-qubit CR Hamiltonian tomography with a target spectator qubit, varying the spectator-target detuning $\Delta_{st}$, frequency of coupler $C_3$ and CR drive amplitude, with a similar pulse sequence sketched in right panel of Fig.~\ref{fig:Fig4}(A). Compared with the perturbation impact from the control spectator qubit, we find that the target spectator qubit affect the CR gate operation more seriously as $\Delta_{st}$ is close to resonance poles, due to the stronger unwanted energy excitations. In fact, this can be understood that a slight jitter of the target qubit frequency either from the static $ZZ$ interaction between $Q_3$ and $Q_4$ or the unwanted energy resonance at $\Delta_{st}=0$ between $\ket{001}$ and $\ket{010}$ as an example, will seriously disturb or even break down the CR gate operation where the target qubit undertakes the main evolution process whereas the control qubit is not directly excited. The detailed dependency of the interaction terms on the $\Delta_{st}$, $J_{34}$ and CR drive amplitude in the target spectator case can be found in Supplementary Materials~\cite{Supplementary}.

\section{Discussions}

So far, we have characterized the three-qubit CR Hamiltonian tomography and find that unwanted energy level resonances are the leading factors for breakdown of the gate evolution. In this section, we take a step further to explore the perturbation impact of spectator qubits on the CR gate fidelity. The outcomes demonstrate that, to yield a high gate fidelity, the qubit frequencies and coupling strength should be deliberately designed to reach a balance between high CR gate fidelity and feasibility of gate operation, particularly in a large superconducting network, where one qubit could be treated as a gate qubit in one network block but practically behaves as a spectator qubit in another.

\begin{figure*}[t]
\includegraphics{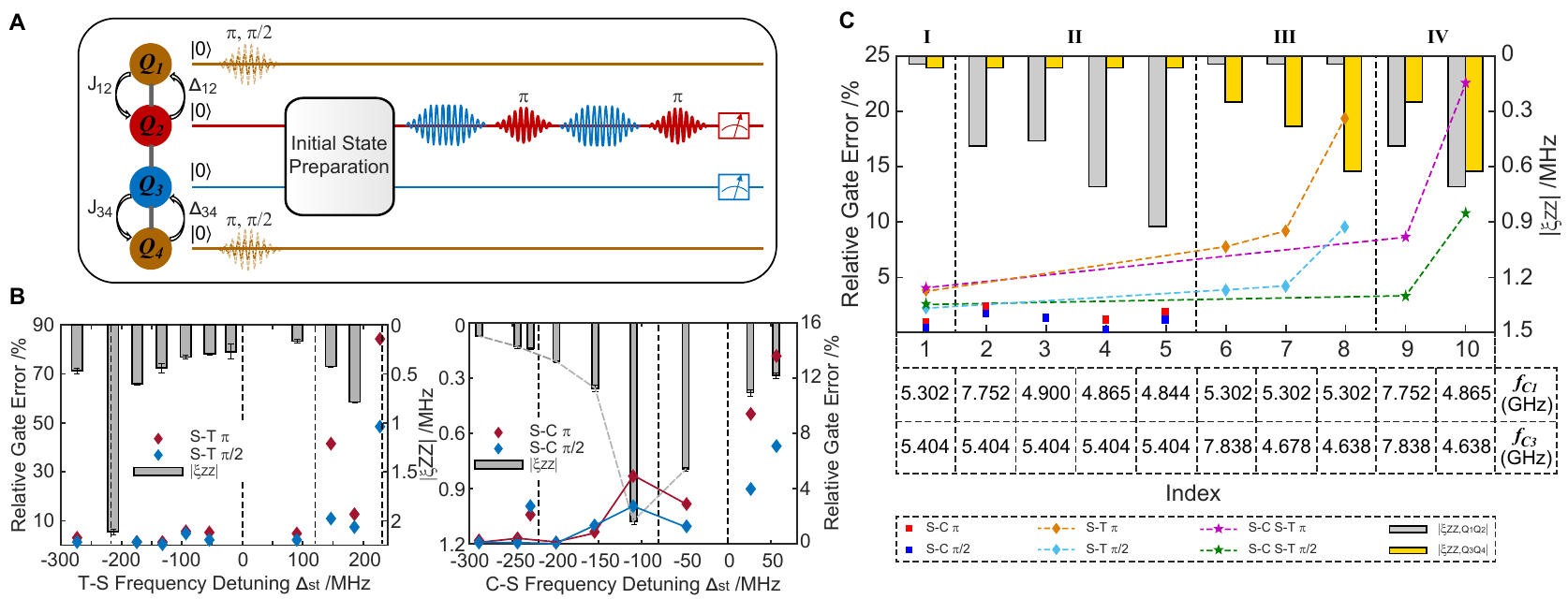}
\caption{(A) The schematic pulse sequence for the exploration of CR gate fidelity with spectator qubits. The qubit frequencies of $Q_2$, $Q_3$ and $C_2$ are biased at $4.426$, $4.289$ and $7.783$ GHz respectively, and $\Delta_{23}$ is fixed at the optimal frequency detuning of $137$ MHz throughout the QPT measurements. We categorize the operation into four regions: I: both couplings ($J_{12}$ and $J_{34}$) off (index 1); II: $J_{12}$ coupling on while $J_{34}$ off (index 2-5); III: $J_{12}$ off while $J_{34}$ on (index 6-8); IV: both $J_{12}$ and $J_{34}$ on (index 9-10). For each operation index, multiple sets of QPT experiments containing idle, $\pi$ or $\pi/2$ pulse on the spectator qubits are used to extract the relative gate error between the control groups and experimental groups. (B) The relative QPT gate error and $ZZ$ interaction (with error bar) vs. the frequency detuning of the spectator $Q_1$ and $Q_4$ to the target qubit $Q_3$, respectively. For right panel, the frequency of $Q_1$ is changed, while biasing the frequency of $C_1$ at $7.654$ GHz (yield a positive coupling between $Q_1$ and $Q_2$) and keeping off the interaction between $Q_3$ and $Q_4$ (via tuning the coupler frequency $C_3$ or adjusting the qubit frequency of $Q_4$ far away from the gate qubits). For left panel, the frequency of $Q_4$ is varied while tuning the frequency of $C_3$ at $7.782$ GHz (yield a positive coupling between $Q_3$ and $Q_4$). The CR gate fidelity is subject not only to the $ZZ$ interaction but also to unwanted energy excitations. The frequency detuning labeled with dotted line refers to the resonance poles which lead to a failure in the CR gate. (C) The relative QPT gate error and $ZZ$ interaction vs. the coupling strength between the spectator qubits and gate qubits. The colored squares, diamonds and stars refer the measured gate error with the indicated pulse sequences applied to the spectator qubits. For example, the red-square point represents a $\pi$ pulse applied on the control-spectator qubit $Q_1$ before the QPT experiment on the gate qubits. The gray and yellow bars show the $ZZ$ interaction between $Q_1$ and $Q_2$, $Q_3$ and $Q_4$, respectively. The qubit frequencies of $Q_1$ and $Q_4$ are tuned at $4.150$ and $4.235$ GHz, respectively. The data points connected by dashed lines highlight the increase of the gate error with the $ZZ$ interaction. }
\label{fig:Fig5}
\end{figure*}

We first investigate the CR gate fidelity susceptible to the frequency detunings between the spectators and the target qubit. We extract the $ZZ$ interaction, between $Q_1$ and $Q_2$, $Q_3$ and $Q_4$, via a Ramsey-type experiment which involves probing the frequency of one qubit with another in either its ground or excited state~\cite{reed2013entanglement,chow2010quantum}. Fig.~\ref{fig:Fig5}(B) shows the relative gate error and $ZZ$ interaction with varying the frequency detuning in both the target spectator case (left panel) and control spectator case (right panel). We observe larger relative gate errors or even failure of the gate near the frequency resonance poles indicated by dashed lines, especially in the target spectator case, revealing that unwanted energy excitations play a major role in degrading the CR gate fidelity. Away from the resonance poles, the gate error, however, relies more on the $ZZ$ interaction, demonstrating certain positive correlations, for instance, among the data points connected by the red, blue and grey straight lines in the control spectator case, where the spectator does not cause deadly impact as the target spectator. 

We then explore the dependency of CR gate fidelity on the coupling strength between the spectators and gate qubits, by modifying the frequency of $C_1$ and $C_3$. The parameters of $Q_2$, $Q_3$ and $C_2$ are chosen according to the optimal condition for isolated two-qubit CR gate operation, while the frequencies of $Q_1$ and $Q_4$, are tuned to the appropriate positions based on the three-qubit Hamiltonian tomography results shown in Fig.~\ref{fig:Fig4}, avoiding the potential resonance poles which break down the gate operation. The $ZZ$ interaction occurs as the coupling between the spectator and gate qubit is on, and its magnitude rises as the increase of the coupling strength, as depicted as color bar shown in Fig.~\ref{fig:Fig5}(C). To probe the perturbation impact in various conditions, we execute multiple sets of QPT experiments for each operation index, selectively applying $\pi$ or $\pi/2$ pulse on either $Q_1$ or $Q_4$, respectively. The experimentally measured QPT results are shown in Fig.~\ref{fig:Fig5}(B) and the corresponding pulse sequence is illustrated in Fig.~\ref{fig:Fig5}(A). The QPT measurement with idle pulse (without pulse) on the spectator qubits (in ground state) sets a control fidelity for each operation index with the particular coupling condition, comparing with the measurements with pulse applied on the spectator qubits (experimental fidelity). The relative gate error in Fig.~\ref{fig:Fig5}, defined as the difference between the experimental fidelity and the control fidelity, reflexes the perturbation impact from the spectator qubits, implying that once the spectator qubit is excited, the $ZZ$ interaction will disturb the CR gate evolution, thus degrading the gate fidelity. As expected, the spectator qubits have almost no perturbation impact on the gate qubit regardless of the operations of the spectator qubits, as the couplings between the gate qubits and spectator qubits are turned off (see index region I). Once the coupling is on, however, the perturbation impact obviously occurs, and the relative gate error increases as the magnitude of $ZZ$ interaction rises. Particularly, the gate qubits are more susceptible to the perturbation impact from the target-spectator qubit (index region III) than that from the control-spectator qubit (index region II). It can be attributed to the fact that the standard echo scheme can only effectively reduce errors caused by control spectators~\cite{Sundaresan_2020}. The perturbation impact becomes more serious, evidenced by the larger relative gate error even up to 22.5\% , when both the couplings are all on (index region IV). For each operation index, $\pi$ pulse on the spectators in general, compared with $\pi/2$ one, yields larger perturbation impact on the gate qubits, demonstrating more notable difference as the coupling between the spectator and target is on, as shown in the operation region of III and IV.     

\section{Conclusions}

In summary, we exploit the flux-controlled tunable coupler to verify the optimal operation condition for constructing the CR gate and provide a guiding principle to improve the CR gate fidelity in large-scale quantum circuits. We here emphasize our main conclusions: (1) We present the first experimental approach to the evaluation of the perturbation impact arising from the spectator qubits. 
The perturbation impact is enhanced on the particular resonance poles where unwanted energy excitations are induced, and the target-spectator qubit leads to more serious degradation of the CR gate fidelity than the control-spectator qubit. (2) We systematically investigate the dependency of gate fidelities on spurious interaction components by tuning the inter-qubit detuning and flux bias on the coupler. The interaction terms rely on the coupling strength, the frequency detuning between the spectator and gate qubits, and the CR drive amplitude. The dominant interaction terms are more pronounced in the resonance pole regions. (3) The three-qubit Hamiltoninan tomography method we develop here can be extended and applied to other multibody systems to extract multi-qubit Hamiltonian interaction terms. Our experimental outcomes will be highly desirable as the CR gate implementation becomes more widely used in large scale superconducting circuits and fault-tolerant quantum computation.

\bigskip
\begin{acknowledgements}
We acknowledge Luyan Sun for sharing JPA fabrication parameters. This work was supported by the National Natural Science Foundation of China under Grant No.11874235, the State's Key Project of Research and Development Plan under Grant No. 2016YFA0301902 and the Tsinghua University Initiative Scientific Research Program.
\end{acknowledgements}

%\bibliographystyle{Zou}
%\bibliography{Citation}

%Control: production of eprint (-1) disabled
%

\end{document}

\title{Supplementary Materials for perturbation impact of spectators on a cross-resonance gate in a tunable coupling superconducting circuit}

\author{T.-Q. Cai}
\affiliation{Center for Quantum Information, Institute for Interdisciplinary Information Sciences, Tsinghua University, Beijing 100084, China}

\author{X.-Y. Han}
\affiliation{Center for Quantum Information, Institute for Interdisciplinary Information Sciences, Tsinghua University, Beijing 100084, China}

\author{Y.-K. Wu}
\affiliation{Center for Quantum Information, Institute for Interdisciplinary Information Sciences, Tsinghua University, Beijing 100084, China}

\author{Y.-L. Ma}
\affiliation{Center for Quantum Information, Institute for Interdisciplinary Information Sciences, Tsinghua University, Beijing 100084, China}

\author{J.-H. Wang}
\affiliation{Center for Quantum Information, Institute for Interdisciplinary Information Sciences, Tsinghua University, Beijing 100084, China}

\author{Z.-L. Wang}
\affiliation{Center for Quantum Information, Institute for Interdisciplinary Information Sciences, Tsinghua University, Beijing 100084, China}

\author{H.-Y Zhang}
\affiliation{Center for Quantum Information, Institute for Interdisciplinary Information Sciences, Tsinghua University, Beijing 100084, China}

\author{H.-Y Wang}
\affiliation{Center for Quantum Information, Institute for Interdisciplinary Information Sciences, Tsinghua University, Beijing 100084, China}

\author{Y.-P. Song}\email{ypsong@mail.tsinghua.edu.cn}
\affiliation{Center for Quantum Information, Institute for Interdisciplinary Information Sciences, Tsinghua University, Beijing 100084, China}

\author{L.-M. Duan}\email{lmduan@tsinghua.edu.cn}
\affiliation{Center for Quantum Information, Institute for Interdisciplinary Information Sciences, Tsinghua University, Beijing 100084, China}

\maketitle

\tableofcontents

\newpage

\section{Hamiltonian Tomography with Spectator Qubits}

\begin{figure*}[hbt]
{\centering\includegraphics[width=1\textwidth]{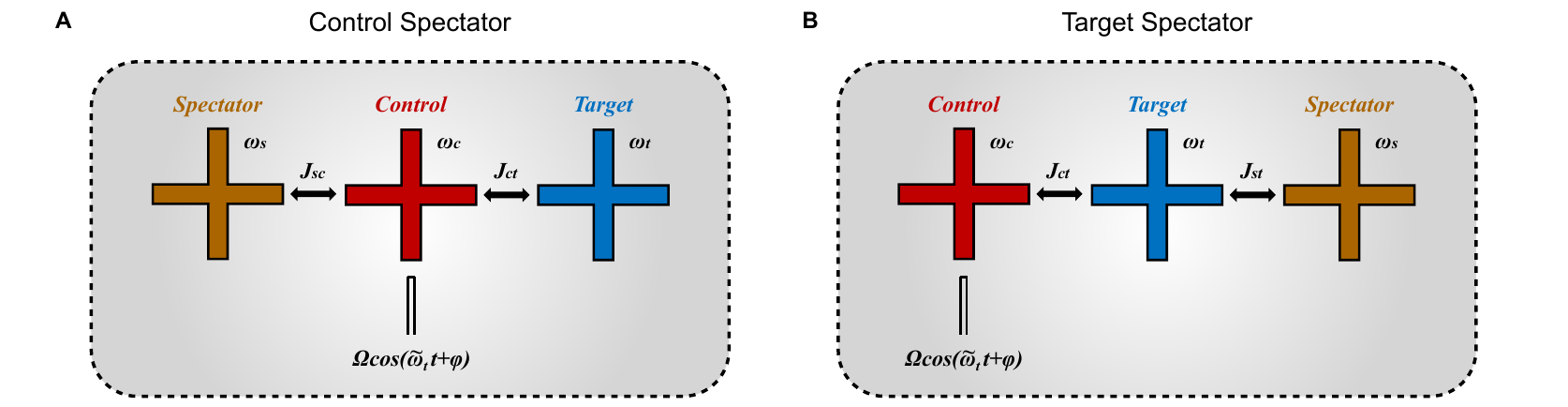}}
\caption{\textbf{Schematic circuit for CR gate with spectator qubits.} 
Schematic circuit for CR gate with (A) a control spectator, and (B) a target spectator. Note that $J_{sc}$, $J_{ct}$ and $J_{st}$ are effective coupling strength between spectator and control, control and target, target and spectator, respectively.}
\label{fig:FigS1}
\end{figure*}

Here, we theoretically derive the three-qubit Hamiltonian tomography with spectator qubits used in our experiment~\cite{sheldon2016procedure,Malekakhlagh_2020}. Without loss of generality, we consider the situation with either a control spectator or a target spectator in the schematic circuit depicted in Fig.~\ref{fig:FigS1}. 

Similar to the effective block-diagonal Hamiltonian model for acquiring CR Hamiltonian, here we consider that the spectator qubit only contributes $\{ I,\, Z \}$ interactions to the effective Hamiltonian owing to the CR pulse scheme, referring to non-excitation of the spectator qubit during the gate operation. Therefore a full CR Hamiltonian can be expressed as (gate operators with a control-spectator as an  example):
\begin{equation}\label{EqS1}
\begin{split}
H=\frac{I \otimes I \otimes A}{2}+\frac{I \otimes Z \otimes B}{2}+\frac{Z \otimes I \otimes C}{2}+\frac{Z \otimes Z \otimes D}{2},
\end{split}
\end{equation}
where $A,\, B,\, C,\, D \in \{X,\, Y,\, Z \}$. According to this Hamiltonian, it is easy to find that only the target qubit involves all the exchange terms due to the off-resonance drive on the control qubit. Expanding the three-qubit CR Hamiltonian with corresponding undetermined coefficients, we then acquire $H=\frac{1}{2}(h_1 IIX + h_2 IIY + h_3 IIZ) + \frac{1}{2}(h_4 IZX + h_5 IZY + h_6 IZZ) + \frac{1}{2}(h_7 ZIX + h_8 ZIY + h_9 ZIZ) + \frac{1}{2}(h_{10} ZZX + h_{11} ZZY + h_{12} ZZZ)$. Projecting the Hamiltonian into the $spectator \otimes control$ subspace with $\ket{00},\, \ket{01},\, \ket{10},\, \ket{11}$ respectively, we can correspondingly restore the evolution of the target qubit as:
\begin{equation}\label{EqS2}
\begin{split}
H_{00}&=\bra{00} H \ket{00}=\frac{1}{2}(h_1+h_4+h_7+h_{10})X+\frac{1}{2}(h_2+h_5+h_8+h_{11})Y+\frac{1}{2}(h_3+h_6+h_9+h_{12})Z, \\
H_{01}&=\bra{01} H \ket{01}=\frac{1}{2}(h_1-h_4+h_7-h_{10})X+\frac{1}{2}(h_2-h_5+h_8-h_{11})Y+\frac{1}{2}(h_3-h_6+h_9-h_{12})Z, \\
H_{10}&=\bra{10} H \ket{10}=\frac{1}{2}(h_1+h_4-h_7-h_{10})X+\frac{1}{2}(h_2+h_5-h_8-h_{11})Y+\frac{1}{2}(h_3+h_6-h_9-h_{12})Z, \\
H_{11}&=\bra{11} H \ket{11}=\frac{1}{2}(h_1-h_4-h_7+h_{10})X+\frac{1}{2}(h_2-h_5-h_8+h_{11})Y+\frac{1}{2}(h_3-h_6-h_9+h_{12})Z.
\end{split}
\end{equation}
Since the Rabi oscillations can be fitted with the Bloch equation as mentioned in the main text, hence the Hamiltonian for target qubit can be represented as:
\begin{equation}\label{EqS3}
\begin{split}
H_{ij}=\frac{1}{2} \Omega^{ij}_x X + \frac{1}{2} \Omega^{ij}_y Y - \frac{1}{2} \Delta^{ij}_z Z,
\end{split}
\end{equation}
where $i,\, j \in \{ \ket{00},\, \ket{01},\, \ket{10},\, \ket{11} \}$~\cite{sheldon2016procedure}. Comparing Eq.~(\ref{EqS2}) with Eq.~(\ref{EqS3}), we can finally calculate the corresponding coefficients in Eq.~(\ref{EqS1}). For instance, $IIX=\frac{1}{8}(\Omega^{00}_x+\Omega^{01}_x+\Omega^{10}_x+\Omega^{11}_x)$, $IZX=\frac{1}{8}(\Omega^{00}_x-\Omega^{01}_x+\Omega^{10}_x-\Omega^{11}_x)$, $ZZX=\frac{1}{8}(\Omega^{00}_x-\Omega^{01}_x-\Omega^{10}_x+\Omega^{11}_x)$, $ZZZ=\frac{1}{8}(-\Delta^{00}+\Delta^{01}+\Delta^{10}-\Delta^{11})$.

\newpage

\section{Mesurement Setup}

Our quantum processor consists of thirteen Xmon qubits with seven Xmon qubits acting as computational qubits and another six Xmon qubits used as tunable couplers~\cite{Devoret_2013,You_2011,Gu_2017,Kelly_2015,Barends_2013,Li_2020,Han_2020}. The coupler qubit is sandwiched between each pair of the computational qubits together with a cross-shaped capacitor, contributing to the total coupling strength between the two computational qubits. Each computational qubit has an independent readout cavity for qubit state measurement. The first four qubits ($Q_1 - Q_4$) share one transmission line and the others ($Q_5 - Q_7$) couple to another transmission line for readout. Fabrication procedure of the sample is similar to that presented in Ref.~\onlinecite{Li_2020}. 

\begin{figure*}[hbt]
{\centering\includegraphics[width=1\textwidth]{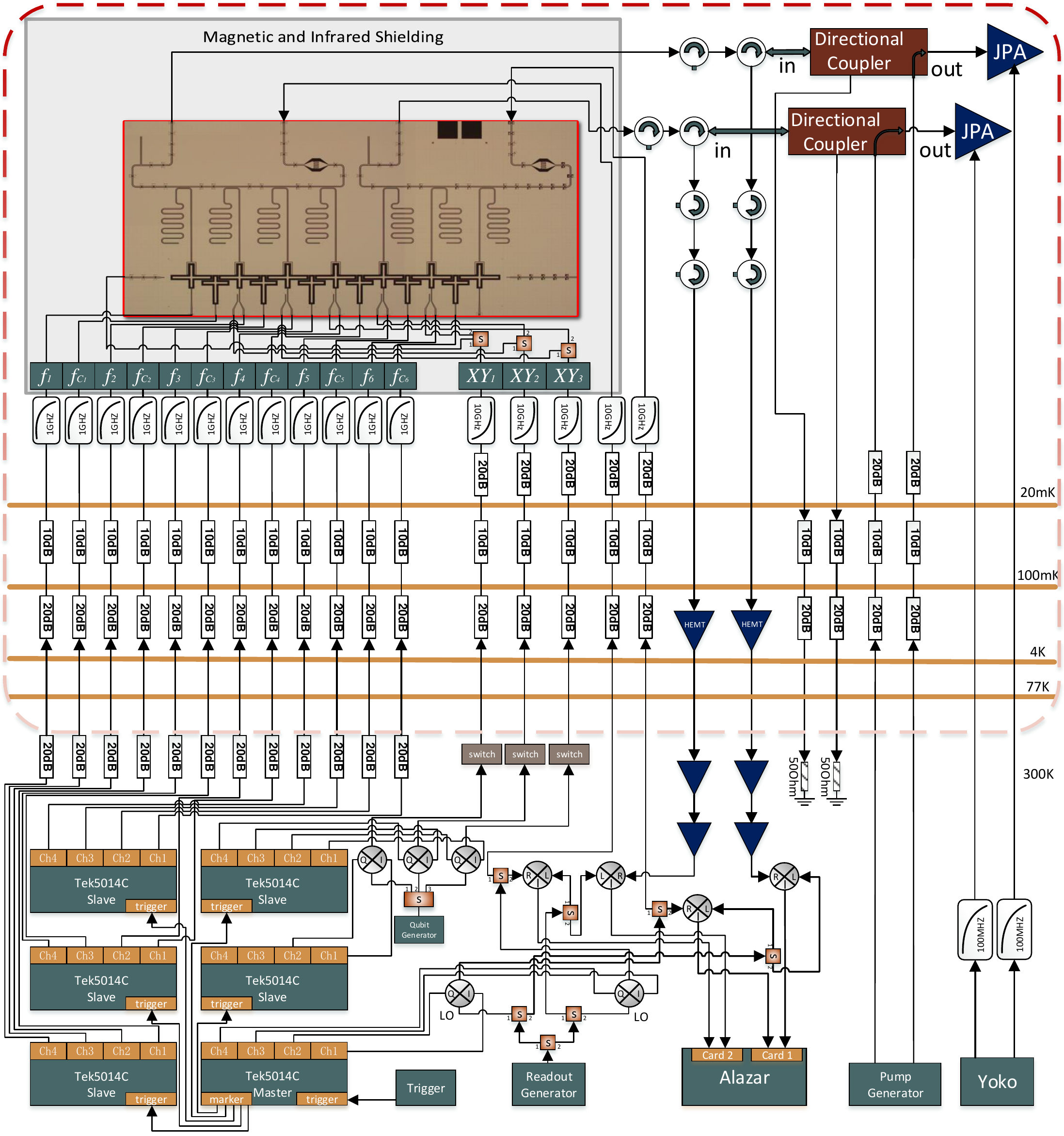}}
\caption{\textbf{Experimental setup.} 
Our measurement circuit contains six AWGs (Tek5014C), four signal generators, and other microwave components. The $XY$ control pulses are generated from a signal generator modulated by the AWG. Flux pulses are generated directly from Tek5014C (with an extra $20$ dB attenuator). Readout signal is amplified by a JPA at the base, a HEMT amplifier at $4$ K and two room-temperature amplifiers, and finally down-converted and digitized by an analog-to-digital converter (ADC).}
\label{fig:FigS2}
\end{figure*}

Our measurement circuitry is depicted in Fig.~\ref{fig:FigS2}. The aluminum sample box is protected via a magnetic shielding and an infrared shielding (shown as grey shaded part), and is loaded in a dilution refrigerator with a base temperature about $10$ mK. Each of computational qubit (except for $Q_7$) and coupler has an independent flux line used for adjusting frequency. For a full manipulation of the device, we use six Arbitrary Waveform Generators (AWGs) (Tek5014C). One AWG is connected to the input-output line for simultaneous readout of the qubits. Another two AWGs, synchronized with the first one, provide three pairs of sideband modulations for $XY$ control. The $XY$ control signals are generated from a single microwave signal generator modulated with different sideband frequencies. This method of control guarantees stable phase differences during the quantum tomography experiments. Due to the limitation of microwave lines in our refrigerator, we combine the $XY$ line for $Q_1$ and $Q_6$, $Q_2$ and $Q_5$, $Q_3$ and $Q_4$ with a cryogenic splitter. Note that manipulation of $Q_7$ and each coupler $C_j \,(j=1 \sim 6)$ could be achieved through the microwave crosstalk. The other AWGs directly generate flux pulses to realize individual $Z$ control of qubits and couplers (with an extra $20$ dB attenuator). A derivative removal adiabatic gate (DRAG) pulse is used for single-qubit rotation and pulse correction to reduce phase error and leakage to higher transmon levels~\cite{Motzoi_2009}. The readout cavity is coupled to a transmission line, which connects to a Josephson parametric amplifier (JPA)~\cite{hatridge2011dispersive,Roy_2012,kamal2009signal,murch2013observing}. The JPA, which is pumped and biased by a signal generator and a voltage source (YOKOGAWA GS210 DC Voltage/Current Source) respectively, has a gain of more than $20$ dB and a bandwidth of about $300$ MHz at $10$ mK. It is used as the first stage of amplification followed by a high-electron mobility transistor amplifier at $4$ K and room-temperature amplifiers, allowing for a high-fidelity simultaneous single-shot readout for all the qubits. In the JPA circuit design, $50$ $\Omega$ impedance matching is applied without any other specific impedance engineering.

\newpage

\section{Device Parameters}

\begin{table*}[ht]
\caption{Qubit parameters. $\omega_r$ is the resonant frequency of the readout cavity for each qubit. Note that the readout cavities of $Q_1 - Q_4$ connect to the same transmission line while the readout cavities of $Q_5 - Q_7$ couple to another one. $\omega_{i,ge,opt}\, (i=1 \sim 7)$ are the maximum resonant frequencies at the sweet spot for qubits, and $\omega_{i,ge,idle}\, (i=1 \sim 7)$ are the idle frequencies for implementing the CR gate with the spectator qubits. $\alpha_i \, (i=1 \sim 7)$ are the anharmonicities of each qubit. $T_1$, $T_{2,opt}$ and $T_{2E,opt}$ are the corresponding energy relaxation time, Ramsey dephasing time and echoed dephasing time for the qubits measured at the sweet spot. $F_{gg,idle}$ ($F_{ee,idle}$) is the typical readout fidelity for detecting each qubit in $\ket{g}$ ($\ket{e}$) when it is prepared in $\ket{g}$ ($\ket{e}$). The outcomes can be used to acquire a calibration matrix for reconstructing qubit readout via the Bayes' rule. $\nu_{i,01,idle},\, \nu_{i,12,idle} (i=2,\, 3)$ are the typical unitless charge operators at the idle frequency defined in the energy-basis representation method used for extracting interaction terms in the CR gate~\cite{Malekakhlagh_2020}.}
\begin{threeparttable}
\begin{tabular}{cp{2.5cm}<{\centering}p{2cm}<{\centering}p{2cm}<{\centering}p{2cm}<{\centering}p{2cm}<{\centering}p{2cm}<{\centering}p{2cm}<{\centering}p{2cm}<{\centering}}
\hline
\hline
 &{$Q_1$} &{$Q_2$} &{$Q_3$} &{$Q_4$} &{$Q_5$} &{$Q_6$} &{$Q_7$} \tabularnewline
\hline
$\omega_r$ (GHz) &{$7.167$} &{$7.185$} &{$7.192$} &{$7.212$} &{$7.191$} &{$7.221$} &{$7.245$} \tabularnewline
$\omega_{i,ge,opt}$ (GHz) &{$4.484$} &{$4.455$} &{$4.393$} &{$4.515$} &{$4.560$} &{$4.383$} &{$4.545$} \tabularnewline
$\omega_{i,ge,idle}$ (GHz) &{$4.150$} &{$4.426$} &{$4.289$} &{$4.235$} &{$\sim$} &{$\sim$} &{$\sim$} \tabularnewline
$\alpha_i/2\pi$ (MHz) &{$-220$} &{$-218$} &{$-218$} &{$-213$} &{$-222$} &{$-220$} &{$-221$} \tabularnewline
$T_1$ ($\mu$s) &{$27.9$} &{$30.7$} &{$29.2$} &{$31.7$} &{$31.0$} &{$26.9$} &{$34.9$} \tabularnewline
$T_{2,opt}$ ($\mu$s) &{$35.0$} &{$26.5$} &{$16.9$} &{$24.6$} &{$33.3$} &{$19.4$} &{$21.7$} \tabularnewline
$T_{2E,opt}$ ($\mu$s) &{$40.3$} &{$32.8$} &{$25.2$} &{$30.0$} &{$37.1$} &{$24.2$} &{$29.3$} \tabularnewline
$F_{gg,idle}$ (\%) &{$94.0$} &{$94.6$} &{$94.5$} &{$93.2$} &{$\sim$} &{$\sim$} &{$\sim$} \tabularnewline
$F_{ee,idle}$ (\%) &{$90.5$} &{$90.8$} &{$91.2$} &{$89.7$} &{$\sim$} &{$\sim$} &{$\sim$} \tabularnewline
$\nu_{i,01,idle}$\tnote{*} &{$\sim$} &{$0.9768$} &{$0.9762$} &{$\sim$} &{$\sim$} &{$\sim$} &{$\sim$} \tabularnewline
$\nu_{i,12,idle}$\tnote{*} &{$\sim$} &{$1.3462$} &{$1.3445$} &{$\sim$} &{$\sim$} &{$\sim$} &{$\sim$} \tabularnewline
\hline
\end{tabular} \vspace{0pt}
\label{Table:TableS1}
\begin{tablenotes}
\footnotesize
\item[*] $\nu_{mn}$ is the unitless charge operators defined in Ref.~\onlinecite{Malekakhlagh_2020} where $\nu_{01}=1-\frac{1}{8}\epsilon-\frac{11}{256}\epsilon^2+O(\epsilon^3)$, $\nu_{12}=(1-\frac{1}{4}\epsilon-\frac{73}{512}\epsilon^2)\sqrt{2}+O(\epsilon^3)$. Here $\epsilon$ can be calculated from qubit frequency and anharmonicity by the equation $[9-4(\frac{\alpha}{\omega})]\epsilon^2+16[1-(\frac{\alpha}{\omega})]\epsilon+64(\frac{\alpha}{\omega})=0, \epsilon>0$.
\end{tablenotes}
\end{threeparttable}
\end{table*}

\begin{table*}[ht]
\caption{Coupler parameters. $\omega_{c,ge,opt}\, (c=1 \sim 6)$ are the maximum resonant frequencies at the sweet spot for each coupler. $\omega_{c,ge,idle}\, (c=1 \sim 6)$ are the idle frequencies of the couplers in the experiment. $\alpha_{c,sim}\, (c=1 \sim 6)$ are the simulated anharmonicities for each coupler. $g_{jk}\, (k=1,\, 2)$ represent the direct coupling strength between the qubit and coupler, and $g_{jd}$ represents the direct coupling between each pair of qubits (`$j$' denotes the order number of the couplers and `$k$' denotes the left or right direct coupling with $k=1$ or $k=2$). Note that coupler spectrum could be probed indirectly through the dispersive shift $\chi_{jk}=\frac{(g_{jk})^2(\alpha_i+\alpha_c)}{2(\Delta_{jk}-\alpha_c)(\Delta_{jk}+\alpha_i)}$ induced by the always-on interaction between the qubit and coupler \cite{Zhu_2013,hatridge2011dispersive}. A detailed measurement method can be found in Ref.~\onlinecite{Li_2020}.}
\begin{tabular}{cp{2.5cm}<{\centering}p{2.4cm}<{\centering}p{2.4cm}<{\centering}p{2.4cm}<{\centering}p{2.4cm}<{\centering}p{2.4cm}<{\centering}p{2.4cm}<{\centering}}
\hline
\hline
 &{$C_1$} &{$C_2$} &{$C_3$} &{$C_4$} &{$C_5$} &{$C_6$} \tabularnewline
\hline
$\omega_{c,ge,opt}$ (GHz) &{$7.752$} &{$7.808$} &{$7.838$} &{$7.798$} &{$7.822$} &{$7.785$} \tabularnewline
$\omega_{c,ge,idle}$ (GHz) &{$5.302$} &{$7.783$} &{$5.404$} &{$\sim$} &{$\sim$} &{$\sim$} \tabularnewline
$\alpha_{c,sim}$ (MHz) &{$254$} &{$254$} &{$254$} &{$254$} &{$254$} &{$254$} \tabularnewline
$g_{j1}$ (MHz) &{$63$} &{$63$} &{$63$} &{$63$} &{$63$} &{$63$} \tabularnewline
$g_{j2}$ (MHz) &{$63$} &{$63$} &{$63$} &{$63$} &{$63$} &{$63$} \tabularnewline
$g_{jd}$ (MHz) &{$\sim$} &{$5.5$} &{$\sim$} &{$\sim$} &{$\sim$} &{$\sim$} \tabularnewline
\hline
\end{tabular} \vspace{-6pt}
\label{Table:TableS2}
\end{table*}

\newpage

\section{Flux-line Crosstalk Calibration}

\begin{figure*}[hbt]
{\centering\includegraphics[width=1\textwidth]{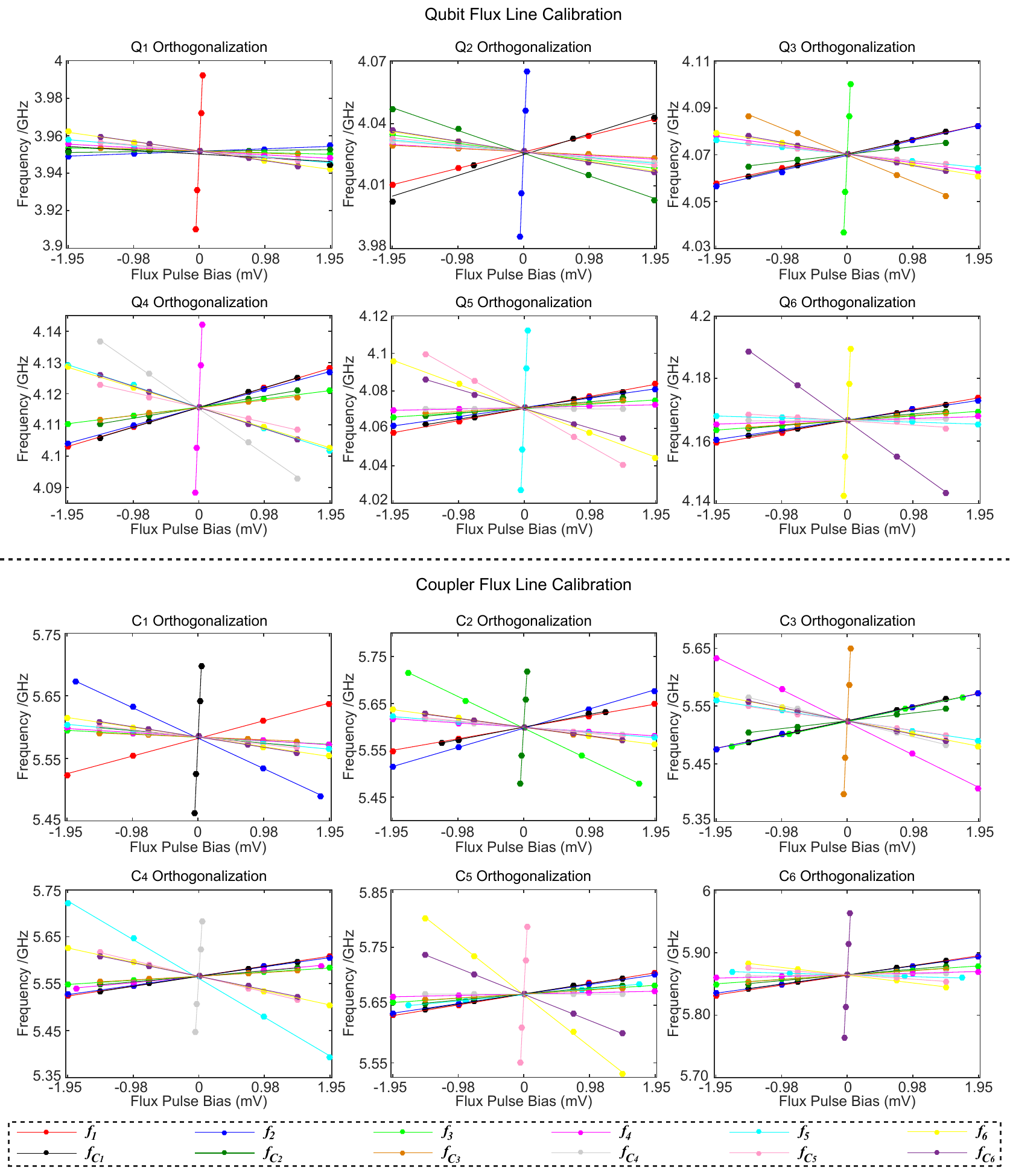}}
\caption{\textbf{Orthogonalization of Flux Line.} 
Orthogonalizations of each flux line for extracting the $Z$-crosstalk calibration matrix for the  computational qubits $Q_1 - Q_6$ and the couplers $C_1 - C_6$. }
\label{fig:FigS3}
\end{figure*}

For purpose of fully exploring the CR gate regime with spectator qubits, we need to accurately and delicately adjust qubit and coupler frequency. We use a fast-flux control to manipulate the qubits and couplers by driving a current into their superconducting quantum interference device (SQUID) loop. However, the return path of the current on each line is not explicitly controlled and, accordingly, there always exists a dc crosstalk between each flux line. Therefore, varying the bias on any individual flux line actually changes all frequencies of the qubit and coupler. Fortunately, since the frequency dependency is approximately linear for a small voltage, similar to our previous work in Ref.~\onlinecite{Li_2020} and Ref.~\onlinecite{Han_2020}, we correct the crosstalk by the orthogonalization of the flux bias lines through multiplying the correction matrix~\cite{reed2013entanglement}. The orthogonalization of each flux line is depicted in Fig.~\ref{fig:FigS3}. The top panel and bottom panel present the flux-line calibration for each computational qubit and coupler, respectively. By fitting the orthogonalizations, we calculate the corresponding Z-crosstalk calibration matrix throughout $Q_1 - Q_6$ and $C_1 - C_6$ as follows: 
\begin{center}
$\widetilde{M}_z$=$M_z^{-1}$=
$\left(\begin{tabular}{ccccccccccccc}
0.9998 & 0.0022 & -0.0017 & -0.0006 & 0.0009 & 0.0013 & 0.0023 & 0.0036 & 0.0039 & 0.0044 & 0.0064 & 0.0067 \\
-0.0126 & 0.9994 & 0.0205 & 0.0044 & 0.0025 & 0.0018 & 0.0028 & 0.0040 & 0.0045 & 0.0053 & 0.0072 & 0.0076 \\
-0.0102 & -0.0130 & 0.9992 & 0.0136 & 0.0053 & 0.0020 & 0.0022 & 0.0032 & 0.0037 & 0.0043 & 0.0064 & 0.0067 \\
-0.0107 & -0.0118 & -0.0181 & 0.9993 & 0.0281& 0.0084 & 0.0041 & 0.0049 & 0.0050 & 0.0059 & 0.0083 & 0.0083 \\
-0.0098 & -0.0105 & -0.0106 & -0.0056 & 0.9994 & 0.0177 & 0.0061 & 0.0049 & 0.0047 & 0.0050 & 0.0076 & 0.0083 \\
-0.0097 & -0.0102 & -0.0100 & -0.0058 & -0.0102 & 0.9995 & 0.0222 & 0.0112 & 0.0073 & 0.0069 & 0.0093 & 0.0098 \\
-0.0119 & -0.0121 & -0.0113 & -0.0071 & -0.0055 & -0.0051 & 0.9995 & 0.0270 & 0.0134 & 0.0095 & 0.0127 & 0.0134 \\
-0.0092 & -0.0092 & -0.0088 & -0.0054 & -0.0041 & -0.0037 & -0.0057 & 0.9997 & 0.0347 & 0.0148 & 0.0137 & 0.0129 \\
-0.0077 & -0.0070 & -0.0061 & -0.0039 & -0.0030 & -0.0028 & -0.0011 & -0.0000 & 0.9998 & 0.0229 & 0.0158 & 0.0128 \\
-0.0080 & -0.0077 & -0.0074 & -0.0048 & -0.0035 & -0.0033 & -0.0013 & -0.0002 & -0.0046 & 0.9999 & 0.0378 & 0.0202 \\
-0.0078 & -0.0071& -0.0069 & -0.0042 & -0.0035 & -0.0035 & -0.0015 & -0.0005 & 0.0012 & 0.0029 & 1.0001 & 0.0319 \\
-0.0076 & -0.0072 & -0.0073 & -0.0047 & -0.0037 & -0.0037 & -0.0015 & -0.0008 & 0.0012 & 0.0034 & 0.0061 & 1.0000 \\
\end{tabular}\right)$\vspace{2pt},
\end{center} 
where $M_z$ presents the matrix elements of qubit frequency response in a basis of $\{ Q_1,\, C_1,\, Q_2,\, C_2,\, \cdots \}$.

\newpage

\section{More Measurement Results for Isolated CR Gate}
\begin{figure*}[hbt]
{\centering\includegraphics[width=1\textwidth]{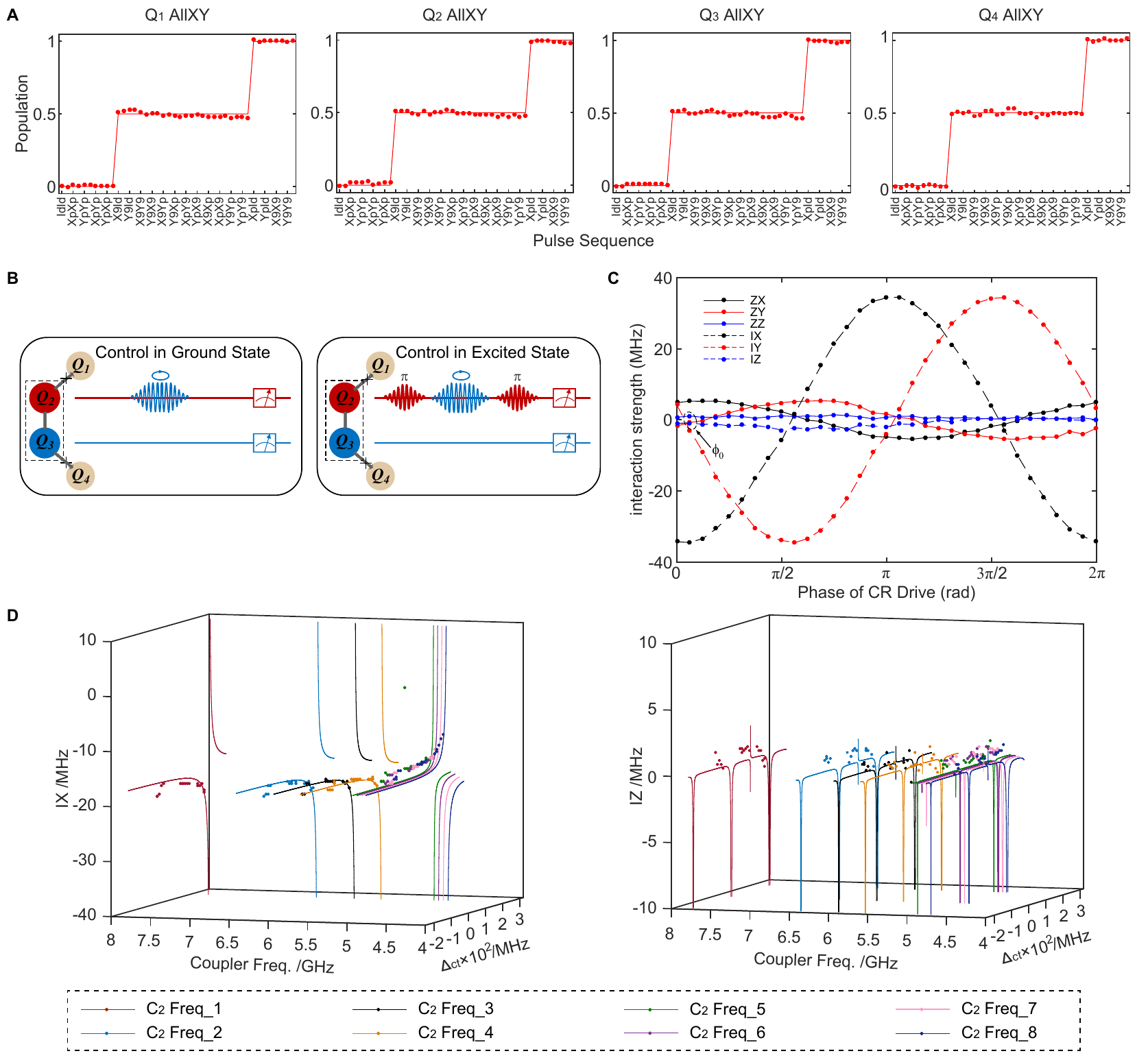}}
\caption{\textbf{Mesurement results for isolated CR gate.} 
(A) $AllXY$ sequence for characterizing single-qubit calibration (`Id' denotes idle pulse, `Xp' denotes $R_x(\pi)$, `Yp' denotes $R_y(\pi)$, `Xp' denotes $R_x(\pi/2)$, `Y9' denotes $R_y(\pi/2)$). Note that each pulse sequence is repeated twice for averaging. (B) Schematic pulse sequence for performing CR Hamiltonian tomography with varying CR drive phase. (C) CR Hamiltonian components vary as a function of the drive phase. $\phi_0$ is the phase at which the $ZX$ component is maximized while the $ZY$ is zero. (D) $IX$ and $IZ$ interactions extracted from CR Hamiltonian tomography (dots) vary with $\Delta_{ct}$ and coupler ($C_2$) frequency with fixed CR drive amplitude $\Omega=18$ MHz. Numerical calculations (solid lines) based on the lowest-order energy basis representation method are illustrated together with the raw data.}
\label{fig:FigS4}
\end{figure*}

In the main text, we fully explore the CR Hamiltonian tomography with varying the frequency detuning $\Delta_{ct}$, effective coupling strength $J_{ct}$ and drive amplitude $\Omega$~\cite{sheldon2016procedure,Malekakhlagh_2020}. Here, we realize the isolated CR gate between $Q_2$ and $Q_3$ by turning off the coupling between the gate qubits and the spectator qubits ($Q_1$ and $Q_2$, $Q_3$ and $Q_4$). The measured $ZZ$ interaction ($\xi_{ZZ,Q_1Q_2}$ and $\xi_{ZZ,Q_3Q_4}$) is lower than tens of kHz. For each data point, the complete single-qubit calibration is automatically performed before executing the CR gate between the gate qubits. Derivative removal adiabatic gate (DRAG) pulse is used for single qubit rotation and pulse collection to reduce the phase error and leakage to higher transmon levels~\cite{Motzoi_2009}. To ensure consistency, the single-qubit operation time is maintained at $40$ ns. Meanwhile, the high quality single-qubit gate is verified via performing $AllXY$ pulse sequence which contains all combinations of one or two single qubit rotations along $x$- and $y$-axes with an angle of $\pi/2$ or $\pi$. Representative results of $AllXY$ for $Q_1 - Q_4$ at the idle frequency are illustrated in Fig.~\ref{fig:FigS4}(A) with the device parameters listed in Table.~\ref{Table:TableS1} and Table.~\ref{Table:TableS2}. 

To further suppress the unwanted CR components like $ZY$ term, we verify an appropriate CR drive phase by measuring the CR Hamiltonian parameters as a function of the drive phase at each date point. The corresponding results at the idle frequency can be found in Fig.~\ref{fig:FigS4}(C) with the pulse sequence shown in Fig.~\ref{fig:FigS4}(B). We find a phase $\phi_0$, indicated by an arrow in the figure, at which the $ZX$ component is maximized while the $ZY$ is zero~\cite{sheldon2016procedure}. We use $\phi_0$ phase of the drive pulse at each data point for our further investigations including extracting interaction terms and measuring CR gate fidelity in an echo scheme.  Here, we plot the measured $IX$ and $IZ$ interactions (dots) in Fig.~\ref{fig:FigS4}(D), as well as the numerical calculations (solid lines), as a function of control-target detuning $\Delta_{ct}$ at eight different flux biases on the coupler with fixed CR drive amplitude $\Omega=18$ MHz. Note that the extraction method for QPT fidelity is similar to that presented in Ref.~\onlinecite{Han_2020}.

\newpage

\newpage

\section{More Measurement Results for CR Gate with Spectator Qubits}
\begin{figure*}[hbt]
{\centering\includegraphics[width=1\textwidth]{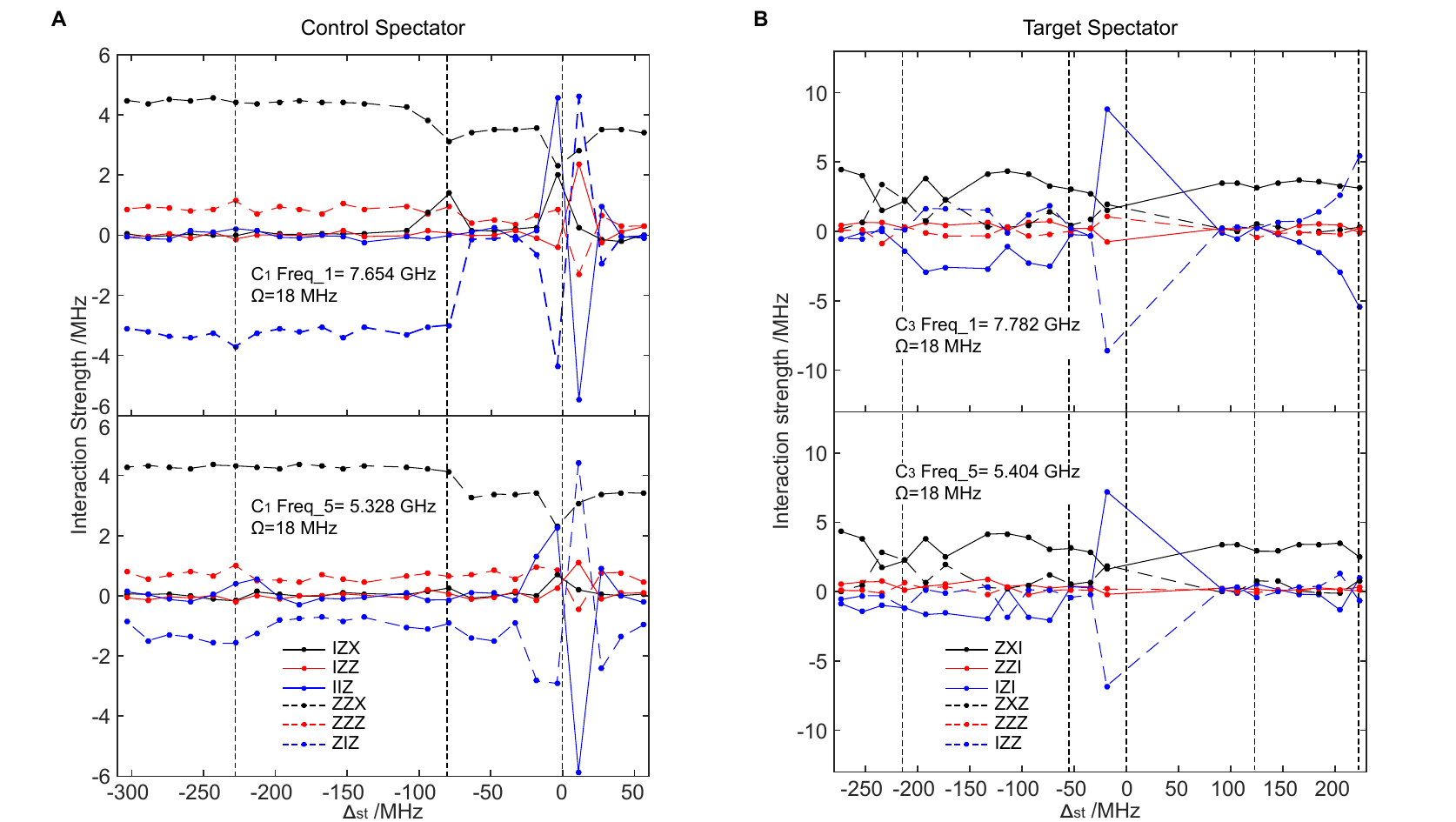}}
\caption{\textbf{Mesurement results for CR gate with spectator qubits.} 
The main interaction terms of the CR gate extracted from three-qubit Hamiltonian tomography with (A) a control spectator qubit, and (B) a target spectator qubit. Specific frequency resonance poles leading to the failure of CR gate regime are marked with dotted lines based on the theoretical calculation.}
\label{fig:FigS5}
\end{figure*}

We have explored and discussed the effects of spectator qubits on the CR gate operation in the main text, revealing that unwanted energy excitation may occur at specific frequency detunings between the spectator qubits and the target qubit~\cite{Malekakhlagh_2020}. Here, we additionally plot more interaction terms as varying the spectator-target detuning $\Delta_{st}$ and coupler frequency with fixed CR drive amplitude $\Omega=18$ MHz for both the control spectator case and the target spectator case, as shown in Fig.~\ref{fig:FigS5}(A)(B). All the possible failure conditions for executing CR gate are marked with dotted line via theoretical calculation. Apparantly, the extracted interaction terms with extreme variations reveal the CR gate is no longer feasible in or near these specific frequency detuning regions, resulting in a failure to fit the data. Note that, since the CR gate evolution is mainly subject to the target qubit, hence the target spectator seems to exert a greater influence on the gate operation, as demonstrated in the figure. Meanwhile, through the comparison between the top panel and bottom panel in both Fig.~\ref{fig:FigS5}(A) and Fig.~\ref{fig:FigS5}(B), we can easily find that stronger couplings between the spectator qubits and gate qubits would induce larger unwanted error terms.

\newpage

%\bibliographystyle{Zou}
%\bibliography{citation_sup}

%Control: production of eprint (-1) disabled
%